\documentclass{article}

\usepackage[english]{babel}
\usepackage{amsmath,amsfonts,amssymb}
\usepackage{graphicx}
\usepackage{float}
\usepackage{braket}
\usepackage{color}
\usepackage[left=2cm,right=2cm,top=2cm,bottom=2cm]{geometry}

\title{Generation of NOON states in waveguide arrays }

\author{Francisco Soto-Eguibar and H\'ector M.  Moya-Cessa\\
	{\small Instituto Nacional de Astrofísica, Óptica y Electrónica, INAOE}\\
	{\small Luis Enrique Erro 1, Santa María Tonantzintla, San Andrés Cholula, Puebla, México, C.P. 72840} }

\begin{document}

\maketitle

\begin{abstract}
	We present a method to generate NOON states with three photons by injecting photons in an array of three waveguides. Conditional measurements project the wave function in a given (desired) state. In passing, we show how the array of three waveguides, that effectively reproduces the interaction of three fields, may be reduced to the interaction of two fields.
\end{abstract}

\section{Introduction}
Entanglement, a distinctive feature of quantum mechanics, has applications in several topics, from its foundations to quantum information theory \cite{jaeger,audretsch}.  The direct interaction between subatomic particles are the most common method to create entangled systems. In optical systems, pairs of photons entangled in polarization may be generated by spontaneous parametric down-conversion. Other methods include entanglement swapping, atomic cascades, quantum dots, the Hong-Ou-Mandel effect, etc. \cite{andres,furusawa}. There are also very strong clues that also in the processes of life, like photosynthesis, entanglement plays an important role \cite{sarovar,lambert,roberto}\\
The generation of nonclassical states has attracted a great deal of attention over the years. Among them, nonclassical states of combined photon pairs such as $W$-states \cite{Dur,Haffner,Leija1} or so-called NOON states \cite{5,6} that, because of their entanglement properties, are particularly useful in  quantum information.\\
It is well known that NOON states can be used to obtain high-precision phase measurements, becoming more and more advantageous as the number of photons grows. Many applications in quantum imaging, quantum information and quantum metrology \cite{7} depend on the availability of entangled photon pairs \cite{epr,sch,jaeger,maxi} that lies at the core of many new applications. These maximally path-entangled multiphoton states may be written in the form 
\begin{equation}
\left| {\mathrm{NOON}} \right\rangle _{a,b}  = \frac{1}{{\sqrt 2
}}\left( {\left| N \right\rangle _a \left| 0 \right\rangle _b  +
\left| 0 \right\rangle _a \left| N \right\rangle _b } \right),
\end{equation}
that represents a superposition of $N$ particles in mode $a$ with zero particles in mode $b$ and zero particles in mode $a$ with $N$ particles in mode $b$.\\
Most schemes to produce NOON states are in the optical regime \cite{6}; however, in the case of cavity fields, it has been shown that NOON states with $N=4$ may be generated \cite{DRodr}, while in ion-laser interactions \cite{Moya} NOON states may be generated for larger $N$'s \cite{Mendez}. Such nonclassical states have also been generated in acoustic wave resonators with high fidelities \cite{Li}.\\
It has been pointed out that NOON states manifest unique coherence properties by showing that they exhibit a periodic transition between spatially bunched and antibunched states when undergo Bloch oscillations, for which the period of bunching/antibunching oscillations is $N$ times faster than the period of the oscillation of the photon density \cite{12}.\\
Bosonic interaction may generate NOON states, and usually the particles involved are photons. NOON states are very valuable in quantum sensing and in quantum metrology. Quantum plasmonic NOON states have been generated in silver nanowires for quantum sensing \cite{Chen}.  Waveguide arrays of three interacting field have been already used to show endurance of quantum coherences due to particle indistinguishability \cite{Montiel}.\\
In this work, a method to generate an entangled NOON state with $N=3$ is presented. Three  interacting quantized fields may evolve from particular initial states and, conditional measurements may be produced, such that they project the wavefunction to a desired (non-classical) state. In Section 2, we present the Hamiltonian of the system and we show that it may be effectively written as the interaction of two quantized fields, that in turn allows us to easily write an evolution operator by using algebraic techniques. In Section 3, we present the result for the special case where we generate NOON states with $N=3$, that  are obtained by the projection of the  state function  via conditional measurements. Finally, Section 4 is devoted to the conclusions.\\

\section{The waveguide array}
Let us consider the interaction between three fields given by the Hamiltonian (we set $\hbar=1$)
\begin{align}\label{ham1}
    \hat{H}=&\omega_0 \hat{a}_0^\dagger \hat{a}_0
    +\omega \left( \hat{a}_1^\dagger \hat{a}_1+ \hat{a}_2^\dagger \hat{a}_2\right)
    +\lambda\left(\hat{a}_1^\dagger \hat{a}_2+\hat{a}_2^\dagger \hat{a}_1 \right)
    \nonumber \\    & 
    + g \left[\hat{a}_0\left( \hat{a}_1^\dagger+ \hat{a}_2^\dagger \right)+\hat{a}_0^\dagger \left(  \hat{a}_1+ \hat{a}_2\right)  \right]     
\end{align}
where the hopping parameters, $\lambda$ and $g$,  denote the rate
at which an excitation may couple from one site to another [see Fig. 1], 
$\left\lbrace \hat{a}_0, \hat{a}_1, \hat{a}_2\right\rbrace $ are the corresponding field annihilation operators and $\left\lbrace \omega_0, \omega \right\rbrace $ are the associated field frequencies.  In order to obtain the Hamiltonian \eqref{ham1} for the three interacting fields, we have used the rotating wave approximation, thus we have assumed that the coupling strength of the fields are much smaller than the field's frequencies; i.e., that $\lambda \ll \omega_0$, $\lambda \ll \omega$, $g \ll \omega$ and $g \ll \omega_0$. The Hamiltonian in Eq. \eqref{ham1} describes the propagation of quantum light in setups of waveguide arrays as the ones depicted in Fig. 1. In Fig. 1(a), we have all the fields interacting, while in Fig. 1(b) a linear waveguide array with interaction to first neighbours.
\begin{figure}[H]
    \centering
    \includegraphics[width=0.7\linewidth]{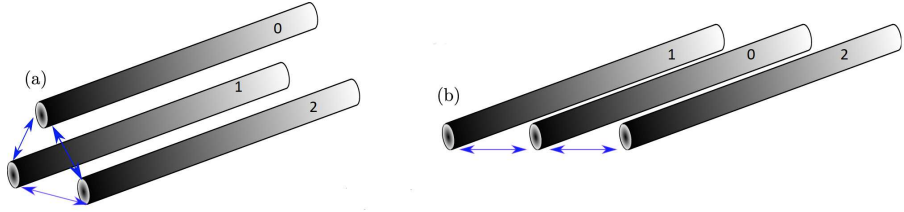}
    \caption{ Possible setups for the Hamiltonian proposed in \eqref{ham1}. (a) is a triangular array while (b) is a linear array. The blue arrows indicate the way in which the waveguides interact: in (a) the three interact with each other, while in (b) the waveguide "1" does not interact with waveguide "2". }
\end{figure}
Our first result is to show that Hamiltonian (2) may be taken to a simpler form, namely the interaction of two quantized fields. In order to achieve this, we define the operators $\hat{A}$ and $\hat{B}$ as
\begin{align}
    \hat{A}=&\frac{\hat{a}_1 +\hat{a}_2}{\sqrt{2}}, 
    \\ 
    \hat{B}=&\frac{\hat{a}_1 -\hat{a}_2}{\sqrt{2}},
\end{align}
such that the Hamiltonian above may be rewritten in the form
\begin{align}\label{ham2}
    \hat{H}=&\omega_0 \hat{a}_0^\dagger \hat{a}_0
    +\left( \omega + \lambda \right)  \hat{A}^\dagger \hat{A}
    +\left( \omega - \lambda \right)  \hat{B}^\dagger \hat{B}
    \\ \nonumber &
    + \sqrt{2} g \left( \hat{A} \hat{a}_0^\dagger + \hat{A}^\dagger \hat{a}_0\right).
\end{align}
These new operators obey the commutation relations 
\begin{equation}
    [\hat{a}_0,\hat{a}_0^\dagger]=1, \qquad [\hat{A},\hat{A}^\dagger]=1, \qquad  [\hat{B},\hat{B}^\dagger]=1,
\end{equation}
and all the other commutators are zero.\\
It is an easy exercise to cast Hamiltonian \eqref{ham2} as
\begin{equation}\label{ham3}
\hat{H}= \Omega _1 \hat{C} +\Omega _2 \hat{J}_z+ \sqrt{2} g
\left(\hat{J}_++\hat{J}_-\right) +  \omega_B  \hat{n}_B,
\end{equation}
with
\begin{equation}\label{080}
\hat{J}_+=\hat{a}_0 \hat{A}^{\dagger },\qquad
\hat{J}_-=\hat{a}_0^{\dagger }\hat{A},\qquad 
\hat{J}_z=\hat{A}^{\dagger } \hat{A}- \hat{a}_0^{\dagger }\hat{a}_0 ,
\end{equation}
and
\begin{equation}\label{090}
\hat{C}= \hat{a}_0^{\dagger } \hat{a}_0 + \hat{A}^{\dagger }\hat{A},
\qquad 
\hat{n}_B= \hat{B}^{\dagger }\hat{B},
\end{equation}
where the new parameters in Eq. \eqref{ham3} are given by
\begin{equation}
\Omega _1=\frac{\omega+\omega_0+\lambda}{2},  \qquad 
\Omega_2=\frac{\omega-\omega_0+\lambda}{2},  \qquad
\omega_B=\omega-\lambda.
\end{equation}
The operators introduced in Eqs. \eqref{080} and \eqref{090} satisfy the following commutation relations
\begin{align}
&\left[\hat{J}_+,\hat{J}_-\right]=\hat{J}_z, \quad
\left[\hat{J}_z,\hat{J}_+\right]=2 \hat{J}_+, \quad
\left[\hat{J}_z,\hat{J}_-\right]=-2\hat{J}_-, 
\end{align}
and all other relevant commutators are zero. \\
The commutators above allow to write the evolution operator $\hat{U}\left(t
\right) =\exp  \left(-i \hat{H} t\right)$ as
\begin{align}
\hat{U}\left(t \right)  =&\exp \left\lbrace  -i \left[ \Omega _1
\hat{C} +\Omega _2 \hat{J}_z+ \sqrt{2} g
\left(\hat{J}_++\hat{J}_-\right) +  \omega_B \hat{n}_B \right] t
\right\rbrace
\nonumber \\
=&\exp  \left(-i t \omega _B  \hat{n}_B  \right) \exp  \left(-i t
\Omega _1  \hat{C}  \right) 
\\ \nonumber
& \times \exp  \left\{-i t\left[\Omega _2
\hat{J}_z+ \sqrt{2} g \left(\hat{J}_++\hat{J}_-\right)
\right]\right\}.
\end{align}
In the Appendix, we prove that because the set $\left\lbrace
\hat{J}_z,\hat{J}_+,\hat{J}_-\right\rbrace $ constitutes an
$\mathrm{su}(1,1)$ algebra, the last part of the evolution
operator can be factorized in the form
\begin{align}\label{opevfac}
& \exp  \left\{-i t  \left[\Omega _2 \hat{J}_z  + \sqrt{2} g
\left(\hat{J}_++\hat{J}_-\right)  \right]\right\}  = \exp \left[-i
f_1\left( t\right) \hat{J}_+\right] 
\\ \nonumber
& \qquad  \times \exp \left[-i f_2\left( t\right)
\hat{J}_z\right] \exp \left[-i f_1\left( t\right) \hat{J}_-\right]
\end{align}
where
\begin{equation}\label{f}
f_1\left( t \right) = i \frac{\Omega _2}{\sqrt{2} g }
-\frac{\sqrt{ 2 g^2+\Omega _2^2}}{\sqrt{2} g } \cot \left(
\theta +t \sqrt{2g^2+\Omega _2^2}\right)
\end{equation}
and
\begin{equation}\label{g}
f_2\left( t \right) =-i \ln \left[   \frac{\sqrt{2} g }{ \sqrt{2g^2+\Omega _2^2}} \sin \left( \theta  + t \sqrt{2g^2+\Omega _2^2} \right)  \right]   ,
\end{equation}
with $\theta= \csc^{-1} \left(   \frac{\sqrt{2} g }{ \sqrt{2g^2+\Omega _2^2}}\right) $ \cite{moya2}.\\
So, finally, we obtain the evolution operator as
\begin{align}\label{evop}
\hat{U}\left(t \right) & =\exp  \left(-i t \omega _B  \hat{n}_B
\right) \exp  \left(-i t \Omega _1  \hat{C}  \right) \exp \left[-i
f_1\left( t\right) \hat{J}_+\right]
\nonumber \\ &
\qquad \times \exp \left[-i f_2\left( t\right)
\hat{J}_z\right] \exp \left[-i f_1\left( t\right) \hat{J}_-\right].
\end{align}
If the total quanta number operator $\hat{N}=\hat{a}_0^\dagger \hat{a}_0+\hat{a}_1^\dagger \hat{a}_1+\hat{a}_2^\dagger \hat{a}_2$ is fixed, it is not difficult to show that $[\hat{H},\hat{N}]=0$, so $\hat{H}$ acts independently in each of the Hilbert spaces with finite number of quanta. It is therefore direct to propose to substitute in the Schr\"odinger equation
\begin{equation}\label{0170}
    i \frac{\partial\ket{\psi}}{\partial t}=\hat{H}\ket{\psi},
\end{equation}
with Hamiltonian \eqref{ham1}, the series solution
\begin{equation}
   \ket{\psi} =\sum_{n_0,n_1,n_2=0}^{N} C_{n_0,n_1,n_2}\left( t \right)  \left|n_0 n_1 n_2\right\rangle
\end{equation}
with the restriction that $n_0+n_1+n_2=N$.\\
In order to find the solution of the Schrödinger equation, \eqref{0170}, with an established initial condition, we apply the evolution operator \eqref{evop} to  that initial state and we obtain the coefficients as function of time. In order to do that, all the operators in the evolution operator \eqref{evop} must be written in terms of the original operators, introduced in the Hamiltonian \eqref{ham1}.

\section{Collapsing through measurements}
We analyse in this section the particular case when we have three photons and the parameters are $\omega_0=\omega$ and $\lambda=0$, reducing the general case depicted in Fig. 1(a) to Fig. 1(b). Notice that the frequencies of the fields and the coupling constants satisfy the hypothesis needed in order to apply the rotating wave approximation. \\
We will consider  as initial state 
\begin{equation}\label{0230}
    \ket{\psi\left( 0 \right) }=\frac{1}{\sqrt{2}}\ket{102}+\frac{1}{\sqrt{2}}\ket{120},
\end{equation}
that obviously is an entangled state, but can be obtained experimentally using the Hong-Ou-Mandel effect \cite{lee,hom}, which allows, in a deterministic way, to produce two photons in one channel.\\
We now apply the corresponding evolution operator \eqref{evop} to  the initial state \eqref{0230} and we obtain the evolved  wave function at time $t$ as
\begin{align}
    \ket{\psi\left(t \right) }&= C_{003}\left(\ket{003}+\ket{030}\right)
     \nonumber \\  &
    +C_{012} \left( \ket{012}+ \ket{021}\right) +C_{102} \left( \ket{102}+\ket{120}\right) 
     \nonumber \\  &
	+C_{111}\ket{111}+C_{201}\left( \ket{201}+\ket{210}\right) +C_{300}\ket{300},
\end{align}
where we are considering $g=0.01$ and all the frequencies equal to one. The ten time dependent coefficients of the wave function are (see Fig. 2, as some of them are equal)
\begin{subequations}\label{coefswf}
	\begin{align}
	C_{003}=&-i \frac{\sqrt{3}}{16}   \left[ 5 \sin \left(\frac{t}{50 \sqrt{2}}
	\right)+\sin \left(\frac{3 t}{50 \sqrt{2}}\right)\right] \exp\left(-3it \right)  \\
	C_{012}=&i \frac{1}{16} \left[ \sin \left(\frac{t}{50 \sqrt{2}}\right)-3\sin
	\left(\frac{3 t}{50 \sqrt{2}}\right)\right]  \exp\left(-3it \right)\\
	C_{102}=&\frac{1}{8 \sqrt{2}}\left[5 \cos \left(\frac{t}{50 \sqrt{2}}\right)+3 \cos\left(\frac{3 t}{50 \sqrt{2}}\right)\right] \exp\left(-3it \right)\\
	C_{111}=&-\frac{3}{2} \sin^2\left(\frac{t}{50 \sqrt{2}}\right) \cos
	\left(\frac{t}{50 \sqrt{2}}\right) \exp\left(-3it \right) \\
	C_{201}=&i \frac{1}{8} \left[\sin \left(\frac{t}{50 \sqrt{2}}\right)-3 \sin \
	\left(\frac{3 t}{50 \sqrt{2}}\right)\right] \exp\left(-3it \right) \\
	C_{300}=&-\sqrt{\frac{3}{2}} \sin ^2\left(\frac{t}{50 \sqrt{2}}\right) \cos \left(\frac{t}{50 \sqrt{2}}\right) \exp\left(-3it \right)
\end{align}
\end{subequations}
In Fig. 2, we present the time behaviour of the absolute value of the ten time dependent coefficients of the three fields wave function. The horizontal axis is time and the vertical axis is the absolute value of the wave function coefficients \eqref{coefswf}.
\begin{figure}[H]
    \centering
    \includegraphics[width=0.7\linewidth]{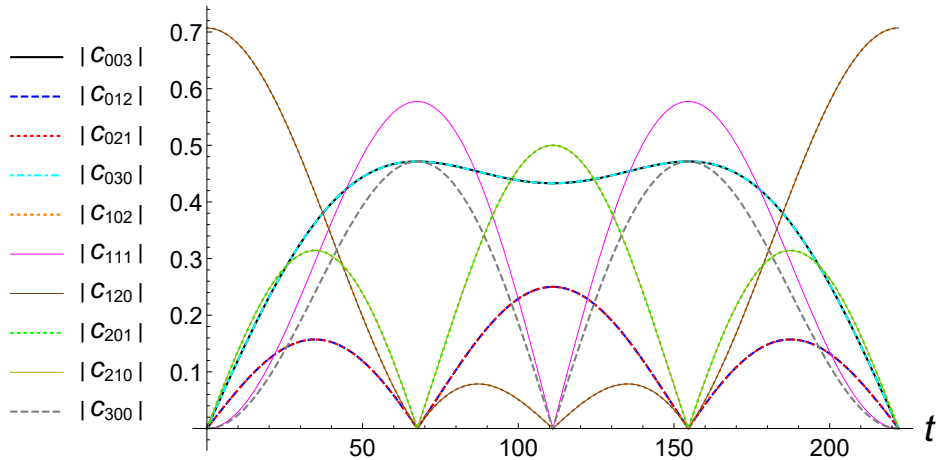}
    \caption{We plot the absolute value of the ten coefficients of the wave function as a function of time for the initial state \eqref{0230} and parameters $g=0.01$ and $\lambda=0$. }
\end{figure}
If a conditional measurement is carried out, obtaining no photons in the "0" waveguide, the wave function collapses to the two fields system state
\begin{align}
\ket{\psi\left(t \right) }= C_{003}\left(\ket{03}+\ket{30}\right) 
+C_{012} \left( \ket{12}+ \ket{21}\right).
\end{align}
In Figure 2, the time evolution of the four coefficients of the
collapsed wave function are shown. It is clear that there are some
times when the coefficient of the state $\ket{12}+\ket{21}$ is
zero, while the coefficient of the state $\ket{03}+\ket{30}$ is
different from zero (in fact, they become maximum, as seen in Fig. 3).
\begin{figure}[H]
    \centering
    \includegraphics[width=0.7\linewidth]{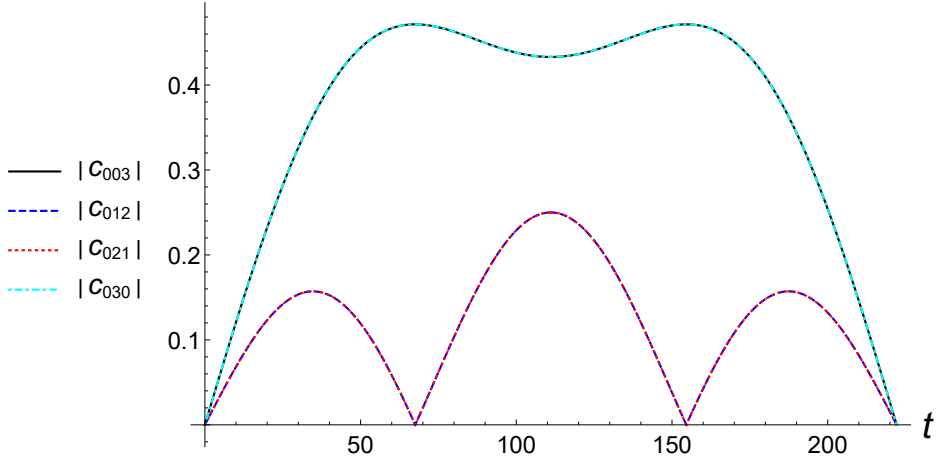}
    \caption{Time evolution of the absolute value of the four coefficients of the collapsed wave function when a conditional measurement have been made obtaining no photons in the field 0 for initial state \eqref{0230} and parameters $g=0.01$ and $\lambda=0$.}
\end{figure}
\noindent Thus, at such times we have that $C_{012}=0$,
i.e., at time $t_1=50\sqrt{2}\arccos\left(\frac{1}{\sqrt{3}} \right)\approx67.6 $
and at time $t_2=50\sqrt{2}\arccos\left(-\frac{1}{\sqrt{3}} \right)\approx 154.6$, the entangled NOON states
\begin{align}\nonumber
    \ket{\psi\left(t_1 \right) }=\ket{\psi\left(t_2 \right) }
    \propto   \ket{03}+\ket{30}
\end{align}
are generated.\\
If another measurement is carried out in the original system and zero photons are obtained in either fields 1 or 2, the wave function collapses to
\begin{align}
\ket{\psi\left(t \right) }=C_{003}  \ket{03}+C_{102}  \ket{12}+C_{201} \ket{21}+C_{300}\ket{30},
\end{align}
where the first component of the kets corresponds to the field 0 and the second to the field 1 or 2, according to the field that it is measured. For times $t_1=50\sqrt{2}\arccos\left(\frac{1}{\sqrt{3}} \right)\approx67.6 $
and  $t_2=50\sqrt{2}\arccos\left(-\frac{1}{\sqrt{3}} \right)\approx 154.6$, we have $C_{102}=C_{201}=0$ and $C_{003}=C_{300}\ne 0$ and we produce the NOON state with $N=3$,
\begin{align}\nonumber
	\ket{\psi\left(t_1 \right) }=\ket{\psi\left(t_2 \right) }
	\propto   \ket{03}+\ket{30}.
\end{align}
Of course, for times not equal to $t_1$ or $t_2$, the $C_{102}$ and $C_{201}$ coefficients are not null simultaneously and we do not get a NOON state. All these arguments are easily visualized in Figure 4, where it is obvious that when both coefficients $C_{102}(t)$ and $C_{201}(t)$ are zero, the other coefficients are equal.
\begin{figure}[H]
    \centering
    \includegraphics[width=0.7\linewidth]{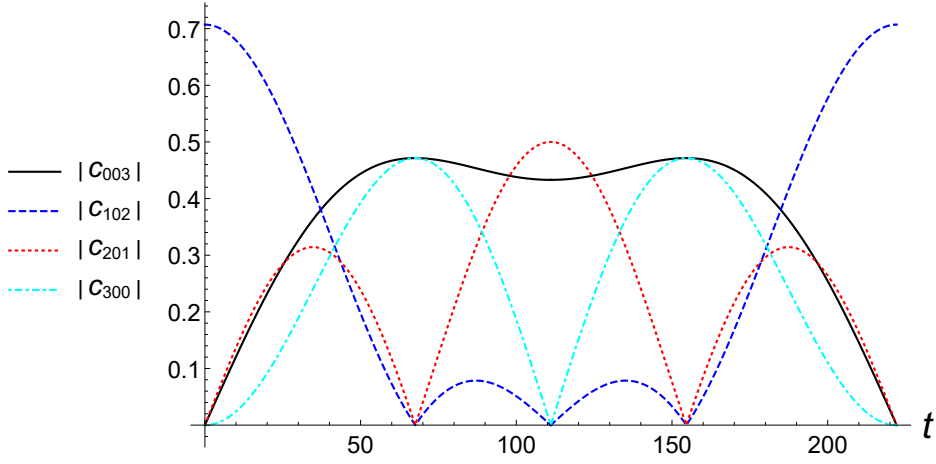}
    \caption{Time evolution of the absolute value of the coefficients of the projected states when zero photons are obtained in fields 1 or 2 for initial state \eqref{0230} and parameters $g=0.01$ and $\lambda=0$. }
\end{figure}

\section{Conclusions}
We have shown a method to generate NOON states in arrays of three waveguides. Our method is robust in the sense that, although it requires conditional measurements, measurements of no-photons in any of the three waveguides generate the NOON state with $N=3$.  In the process, we have shown how the interaction of three fields may be effectively reduced to the interaction of two fields.

\section{Appendix}
In this Appendix, we outline the steps followed to factorize the operator in \eqref{opevfac}. We define
\begin{align}\label{ap010}
&\hat{U}_2\left( t \right)=\exp\left\lbrace -i t \left[ \Omega_2 \hat{J}_z+\sqrt{2} g \left( \hat{J}_+ + \hat{J}_-\right) \right] \right\rbrace  \nonumber \\ &
=\exp\left[-i f_1\left( t \right) \hat{J}_+  \right] 
\exp\left[-i f_2\left( t \right) \hat{J}_z  \right]
\exp\left[-i f_3\left( t \right) \hat{J}_-  \right].
\end{align} 
Differentiating the first line of the previous equation with respect to $t$, we get
\begin{equation}\label{ap020}
\frac{d\hat{U}_2\left( t \right)}{dt}=
-i \left[ \Omega_2 \hat{J}_z+\sqrt{2} g \left( \hat{J}_+ + \hat{J}_-\right) \right] 
\hat{U}_2\left( t \right).
\end{equation} 
Differentiating the second line of Eq \eqref{ap010} with respect to $t$, introducing three times the identity operator written as $\hat{I}=\exp(i \hat{O})\exp(-i \hat{O})$, being the operator $\hat{O}$ a hermitian operator, and collecting terms, we arrive to
\begin{align}\label{ap030}
&\frac{d\hat{U}_2\left( t \right)}{dt}=
-i \left[ \frac{df_1}{dt} \hat{J}_+ 
+ \frac{df_2}{dt} \exp(-if_1\hat{J}_+)\hat{J}_z \exp(if_1\hat{J}_+)
\right.  \nonumber \\  &   \left. 
+ \frac{df_3}{dt} \exp(-if_1\hat{J}_+)\exp(-if_2\hat{J}_z)\hat{J}_-\exp(if_2\hat{J}_z) \exp(if_1\hat{J}_+)\right] 
\nonumber \\ &  \times
\hat{U}_2\left( t \right),
\end{align} 
where, for simplicity, we have dropped the time dependence of the functions.\\
By using the fact that the set $\left\lbrace
\hat{J}_z,\hat{J}_+,\hat{J}_-\right\rbrace $ constitutes a
$\mathrm{su}(1,1)$ algebra, we can easily prove that
\begin{subequations}
	\begin{align}
	&\exp(-if_1\hat{J}_+)\hat{J}_z \exp(if_1\hat{J}_+)=\hat{J}_z + 2 i f_1 \hat{J}_+, \\
	&\exp(-if_2\hat{J}_z)\hat{J}_-\exp(if_2\hat{J}_z)=\exp(2if_2)\hat{J}_-,\\
	&\exp(-if_1\hat{J}_+)\hat{J}_- \exp(if_1\hat{J}_+)=\hat{J}_- - i f_1 \hat{J}_z + f_1^2 \hat{J}_+.
	\end{align}
\end{subequations}
Now, by substituting these relations in \eqref{ap030}, we may write
\begin{align}\label{ap060}
\frac{d\hat{U}_2\left( t \right)}{dt}=&
-i \left[ \frac{df_1}{dt} \hat{J}_+ 
+ \frac{df_2}{dt}\left( \hat{J}_z+2 i f_1 \hat{J}_+ \right)
\right.  \nonumber \\  &   \left. 
+ \frac{df_3}{dt} \exp(2if_2) \left( \hat{J}_- - i f_1 \hat{J}_z+ f_1^2 \hat{J}_+\right) 
\right] 
\hat{U}_2\left( t \right).
\end{align}  
By equating \eqref{ap020} and \eqref{ap060}, doing some algebra and using the linear independence of the operators, we obtain the system of coupled ordinary differential equations
\begin{subequations}\label{ap0370}
	\begin{align}
	& \Omega_2-\frac{df_2}{dt}+ i f \frac{df_3}{dt} \exp\left( 2 i f_2 \right)=0,\\
	& \sqrt{2} g-\frac{df_1}{dt}-2if_1\frac{df_2}{dt}-f_1^2\frac{df_3}{dt}\exp\left( 2 i f_2 \right)=0,\\
    & \sqrt{2} g-\frac{df_3}{dt}\exp\left( -2 i f_2 \right)=0,
	\end{align}
\end{subequations}
with the obvious initial conditions $f_1(0)=f_2(0)=f_3(0)=0$.\\
The solution of the system \eqref{ap0370}, with its corresponding initial conditions, are given by Eqs. \eqref{f} and \eqref{g}, and $f_3(t)=f_1(t)$.

\end{document}